\documentclass[a4paper]{jpconf}
\usepackage{graphicx}
\begin{document}
\title{Alpha-induced reactions for the astrophysical p-process: the case of $^{151}$Eu}

\author{Gy~Gy\"urky$^1$, Z~Elekes$^1$, J~Farkas$^1$, Zs~F\"ul\"op$^1$, G G~Kiss$^1$, E~Somorjai$^1$, T~Sz\"ucs$^1$, R~T~G\"uray$^2$, N~\"Ozkan$^2$, C~Yal\c c\i n$^2$ and T~Rauscher$^3$}

\address{$^1$ Institute of Nuclear Research (ATOMKI), H-4001 Debrecen, POB.51., Hungary}
\address{$^2$ Kocaeli University, Department of Physics, TR-41380 Umuttepe, Kocaeli, Turkey }
\address{$^3$ University of Basel, Department of Physics, CH-4056 Basel, Switzerland}
 
\ead{gyurky@atomki.hu}

\begin{abstract}
The cross sections of $^{151}$Eu($\alpha,\gamma)^{155}$Tb and $^{151}$Eu($\alpha$,n)$^{154}$Tb reactions have been measured with the activation method. Some aspects of the measurement are presented here to illustrate the requirements of experimental techniques needed to obtain nuclear data for the astrophysical p-process nucleosynthesis. Preliminary cross section results are also presented and compared with the predictions of statistical model calculations.
\end{abstract}

\section{Introduction}

\begin{figure}
\centering 
\resizebox{0.8\textwidth}{!}{\rotatebox{270}{\includegraphics{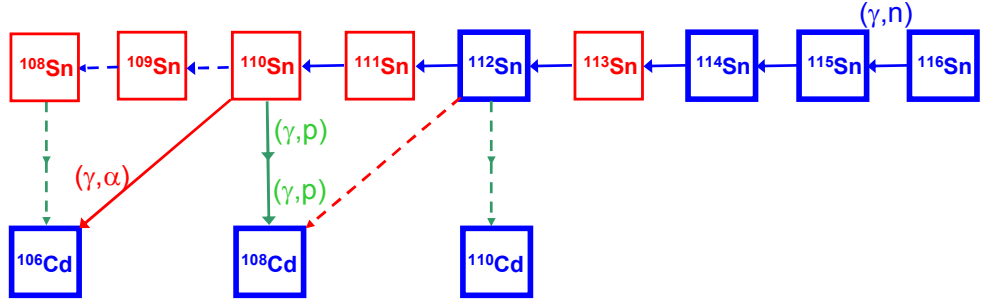}}}
\caption{\label{fig:pprocflow} (Color online) Path of the $\gamma$-process around the Tin isotopes. Thick blue and thin red boxes indicate stable and unstable isotopes, respectively. The main reaction flow according to \cite{gyu06} is shown by solid arrows while dashed arrows show weaker branches of the path. Not all possible reactions are indicated, the figure is intended only to show the different reactions contributing to the synthesis of p-nuclei.}
\end{figure}

The chemical elements with atomic number higher than about 30 are very rare in nature. In the Solar System these trans-iron elements represent only roughly 1 atom in every 10 million. The synthesis of about 99\,\% of the nuclei in this mass region can be explained by neutron capture reactions in the astrophysical s- and r-processes under various astrophysical conditions \cite{bus99,arn07}. There are, however, more than 30 isotopes on the proton rich side of the nuclear chart which can be synthesized by none of these two processes. There are still many open questions regarding the synthesis of these so called p-nuclei. Their production mechanism is the astrophysical p-process which may involve several different sub-processes \cite{arn03}. The most important mechanism is thought to be the $\gamma$-process which proceeds by $\gamma$-induced reactions starting from  pre-existing s- or r-process seed nuclei. Consecutive ($\gamma$,n) reactions drive the material towards the neutron-deficient region, though charged-particle emitting ($\gamma,\alpha$) and ($\gamma$,p) reactions also contribute to the process. Figure \ref{fig:pprocflow} schematically shows the path of the $\gamma$-process in the case of the Tin isotopic chain.

Modelling of the p-process in the whole relevant mass region involves a reaction network of thousands of reactions. The aim of a p-process model calculation is to reproduce the p-isotope abundances observed in the Solar System, however, even the state-of-the-art models fail to give a satisfactory reproduction of the measured abundances (see e.g. \cite{rap06}). The models use various input parameters. On one hand the astrophysical conditions where the process takes place must be known. This involves the knowledge of the stellar composition (seed abundances), temperature, time scale, etc. The presently favoured site for the $\gamma$-process is the O/Ne layers of massive stars in explosive or pre-explosive phase where temperatures of typically 3\,GK are reached for a period of the order of one second \cite{rau02}. On the other hand, the models require the knowledge of different nuclear physics parameters. The masses and decay half-lives of the (mainly proton-rich) isotopes involved in the reaction network are relatively well known. The reaction rates of the dominant $\gamma$-induced reactions are, however, much more uncertain. Since the networks include thousands of reactions on mainly unstable nuclei, the calculations have to rely on theoretical reaction rate calculations obtained typically from Hauser-Feshbach statistical models. The calculated p-isotope abundances are very sensitive to the variation of the reaction rates (see e.g. \cite{rau06}), therefore the experimental study of the reactions relevant for a $\gamma$-process network is important to test the theoretical reaction rates.

\section{Cross section measurement for the $\gamma$-process}

As discussed above, $\gamma$-induced reactions play a key role in the synthesis of p-nuclei. The measurement of $\gamma$-induced reaction cross sections is technically challenging, especially in the case of charged particle emitting ($\gamma,\alpha$) and ($\gamma$,p) reactions. Recently, there has been a significant development in this field (see e.g. \cite{has09} and references therein). However, due to the huge difference
between the reaction rate measured in the laboratory and the actual rate in a stellar plasma (stellar enhancement effect, for details see \cite{moh07}), the experimental $\gamma$-induced reaction cross section data cannot be used directly in a nucleosynthesis model.

The experimental study of radiative capture reactions provides an alternative way to investigate the reactions relevant for the $\gamma$-process, since the rate of the $\gamma$-induced reactions can be inferred from the rate of the inverse capture reaction using the detailed balance theorem. Moreover, the stellar enhancement effect is much less pronounced in the case of capture reactions. While the neutron capture cross sections are rather well known experimentally, at least along the line of stability \cite{bao00}, there are only very few ($\alpha,\gamma$) and -- compared to the size of the reaction network -- a limited number of (p,$\gamma$) cross sections measured in the relevant mass and energy range leaving the statistical model calculations for these reactions largely untested. 

\begin{table}[h]
\caption{\label{tab:measuredisotopes} List of A\,$\geq$\,70 isotopes for which experimental proton or alpha capture reactions cross sections are available}
\begin{tabular}{c|c}
\hline
(p,$\gamma$) & ($\alpha,\gamma$)\\
\hline
\vspace{-3mm}\\
$^{74,76,77}$Se, $^{84,86,87,88}$Sr, $^{89}$Y, $^{90,96}$Zr, $^{93}$Nb, $^{92,94,95,98}$Mo,&$^{70}$Ge,\,$^{96}$Ru,\,$^{106}$Cd,\,$^{112,117}$Sn,\\
$^{96,98,99,100,104}$Ru,\,$^{102,104,105,106}$Pd,\,$^{106,108}$Cd,\,$^{112,114,116,119}$Sn,\,$^{120}$Te&$^{113}$In, $^{144}$Sm, $^{197}$Au\\
\hline
\end{tabular}
\end{table}

Table \ref{tab:measuredisotopes} lists those isotopes with mass number higher than 70 for which experimental proton or alpha capture reactions cross sections are available. It is apparent that in the case of ($\alpha,\gamma$) reactions the available experimental database is very limited especially in the heavier mass region where the p-process models show the highest sensitivity to the ($\alpha,\gamma$) reaction rates. Therefore it is highly needed to extend the experimental database of ($\alpha,\gamma$) cross sections to the heavier mass region. 

\section{Experimental methods for charged particle capture cross section measurements}

The conventional method for capture cross section measurements is the in-beam $\gamma$-spectroscopy. In the mass and energy range relevant for the $\gamma$-process, however, severe problems are encountered in the application of this technique. As it is usual in nuclear astrophysics, low cross sections from millibarn down to the nanobarn range must be measured and the beam-induced background produced on low Z target impurities may hinder the cross section measurement. Moreover, the compound nucleus created in a capture process is typically excited to an energy range where the nuclear energy level density is very high. This results in a complicated decay scheme involving many transitions. All these transitions must be measured with their complete angular distributions which requires huge experimental effort (see e.g. \cite{gal03,has09b}). With the application of a 4$\pi$ summing crystal some of these problems can be avoided \cite{spy07}, but none of the in-beam $\gamma$-spectroscopy methods succeeded so far to reach the heavier (A $>$ 110) mass region of the p-nuclei.

The activation method has been proved to be a very fruitful technique for the determination of cross sections relevant for the $\gamma$-process. Most of the isotopes listed in table \ref{tab:measuredisotopes} have been studied with activation. This method is applicable only if the reaction product of the capture leads to a radioactive isotope with convenient half-life and strong enough decay signature (like a high intensity $\gamma$-radiation). Most of the problems encountered in in-beam $\gamma$-spectroscopy can be avoided in an activation experiment. Since the off-line measurement of the induced activity leads to the determination of the total number of capture reactions, there is no need to consider individual $\gamma$-transitions, angular distributions, etc. The activation technique is also less sensitive to target impurities and in certain cases more isotopes of a given element can be studied simultaneously. 

In the present work the cross section of the $^{151}$Eu($\alpha,\gamma)^{155}$Tb and $^{151}$Eu($\alpha$,n)$^{154}$Tb reactions have been measured with the activation method. (The study of ($\alpha$,n) reactions -- although not directly relevant to the p-process -- provides additional information to the investigation of different optical potentials, see below.) In the next section some details of the measurement are given emphasizing the special requirements of an activation experiment.

\section{Cross section measurement of $^{151}$Eu($\alpha,\gamma)^{155}$Tb and $^{151}$Eu($\alpha$,n)$^{154}$Tb reactions}

As can be seen in table \ref{tab:measuredisotopes}, the only isotope for which $\alpha$-capture cross section measurement has been carried out near the A\,=\,150 mass region is $^{144}$Sm \cite{som98}. For this isotope it was found, that the statistical model calculations using different $\alpha$-nucleus optical potentials give extremely different cross section results. One can choose the best optical model parametrization for a given reaction only if experimental data are available for comparison. Therefore, it is important to collect more experimental data in this mass region to have a larger experimental database to check the performance of different global optical potentials. Our next step towards this goal is the study of $^{151}$Eu. 

\begin{figure}
\centering 
\resizebox{0.6\textwidth}{!}{\rotatebox{270}{\includegraphics{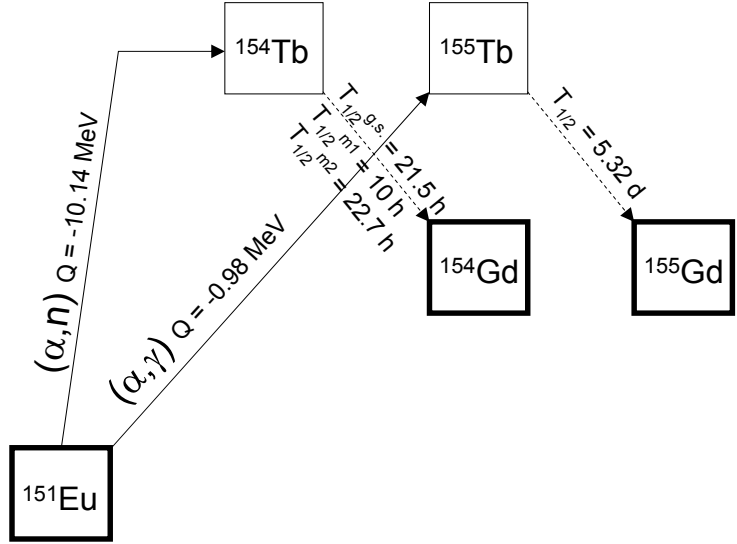}}}
\caption{\label{fig:reaction} The studied reactions and the decay of the reaction products.}
\end{figure}

Alpha capture on $^{151}$Eu (Q\,=\,-1.98\,MeV) leads to $^{155}$Tb which decays to $^{155}$Gd by electron capture with a half-life of 5.32\,d. The decay is followed by a few relatively strong $\gamma$-radiations which can be used for the cross section measurement. Moreover, the ($\alpha$,n) reaction on $^{151}$Eu (Q\,=\,-10.14\,MeV) also produces a radioactive residual nucleus, $^{154}$Tb, so the $^{151}$Eu($\alpha$,n) cross section can also be measured by activation. The decay scheme is, however, more complicated. Besides its ground state decaying to $^{154}$Gd with 21.5\,h half-life, it has two long-lived isomeric states with half-lives of 9.4\,h and 22.7\,h decaying by $\beta^+$, electron capture and/or internal transition. Fortunately all these different decays are followed by many strong $\gamma$-radiations, a few of which can be attributed exclusively to the decay of a given state. Thus, partial cross sections to the ground and the two isomeric states can be determined separately and their sum gives the total  ($\alpha$,n) cross section. Figure \ref{fig:reaction} shows the relevant part of the chart of nuclides with the studied reactions and the decay of the reaction product.

It should be noted that the precision of the decay parameters (half-life, $\gamma$-intensity) directly influences the precision of the half-life determination. In the case of the m1 isomer in $^{154}$Tb, the half-life has a relatively large uncertainty (9.4\,$\pm$\,0.4\,h) and ambiguous values can be found in the literature. Therefore, we have performed a high precision experiment to determine this half-life value. We have obtained T$_{1/2}$\,=\,9.994\,$\pm$\,0.039\,h \cite{gyu09}. With this measurement we reduced the uncertainty of this half-life by one order of magnitude and avoided one possible source of systematic uncertainty in the $^{151}$Eu($\alpha$,n)$^{154}$Tb cross section determination.

\subsection{Experimental technique}

Targets have been prepared by evaporating Eu$_2$O$_3$ enriched to 99.2\,\% in $^{151}$Eu onto thin Al foil. Although the natural abundance of $^{151}$Eu is relatively high (47.8\,\%), enriched target material was necessary to avoid the high background coming from the decay of $^{156}$Tb, the ($\alpha$,n) reaction product of the heavier Eu isotope. $^{156}$Tb has a half-life of 5.4\,d, similar to that of $^{155}$Tb and the $^{153}$Eu($\alpha$,n) cross section is much higher than that of $^{151}$Eu($\alpha,\gamma$). Test experiments with natural isotopic composition targets showed that in the presence of the $^{156}$Tb decay, the weak $\gamma$-radiation from $^{155}$Tb is not measurable.

The number of target atoms enters directly to the calculation of the cross section. Using a thin Al foil as target backing, it was possible to obtain the number of target atoms by measuring the weight of the Al foil before and after the evaporation. It is also important that the target remains stable under $\alpha$-bombardment. The target stability can be monitored by detecting the backscattered $\alpha$-particles from the target continuously during the irradiation. For the detection of the backscattered particles, low Z backing material is preferred in order to have a clear separation of the scattering events from Eu and from the backing. Thus, Al was an ideal choice.

The irradiations have been carried out at the cyclotron accelerator of ATOMKI. The typical beam current was 1\,--\,1.5\,p$\mu$A and based on the backscattering spectra, no target deterioration was observed with this beam intensity. The energy range between 11.5 and 17.5\,MeV was covered with about 0.5\,MeV steps. Owing to the relatively short half-life of the reaction products, the targets could be used more than once at different energies. At some energies the irradiations were carried out with two different targets and consistent results were obtained. The length of the irradiations varied between 5 and 24 hours.

\begin{figure}
\centering 
\resizebox{0.66\textwidth}{!}{\rotatebox{270}{\includegraphics{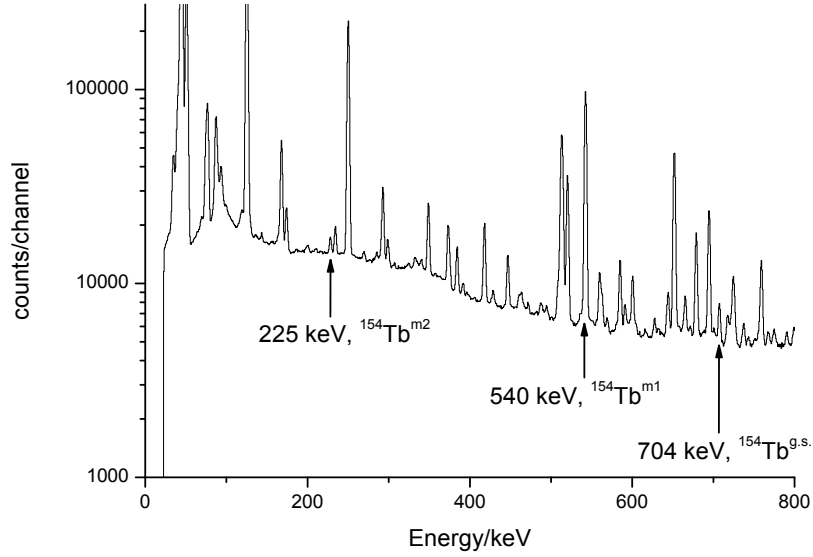}}}
\caption{\label{fig:spectrum} Gamma-spectrum measured for one hour on a target irradiated by 17\,MeV alphas. The strongest peaks uniquely assigned to a given state in $^{154}$Tb are marked with arrows.}
\end{figure}

The induced $\gamma$-activity has been measured with a calibrated 40\,\% relative efficiency HPGe detector equipped with lead shielding. The countings were started typically one hour after the end of the irradiation and lasted for at least 12 hours. In this period the $\gamma$-spectra were dominated by the peaks from the $^{151}$Eu($\alpha$,n)$^{154}$Tb and the $\gamma$-lines from the $^{151}$Eu($\alpha,\gamma)^{155}$Tb reaction were buried under the Compton background. Therefore, a repeated $\gamma$-counting was carried out about one week after the irradiation. At this time the shorter-lived $^{154}$Tb isotope was essentially absent and $^{155}$Tb $\gamma$-peaks could be observed on a strongly reduced background. As an example, figure \ref{fig:spectrum} shows the measured $\gamma$-spectrum directly after the irradiation where many peaks from the decay of the three states of $^{154}$Tb can be seen.

\subsection{Preliminary results}

\begin{figure}
\centering 
\resizebox{0.8\textwidth}{!}{\rotatebox{270}{\includegraphics{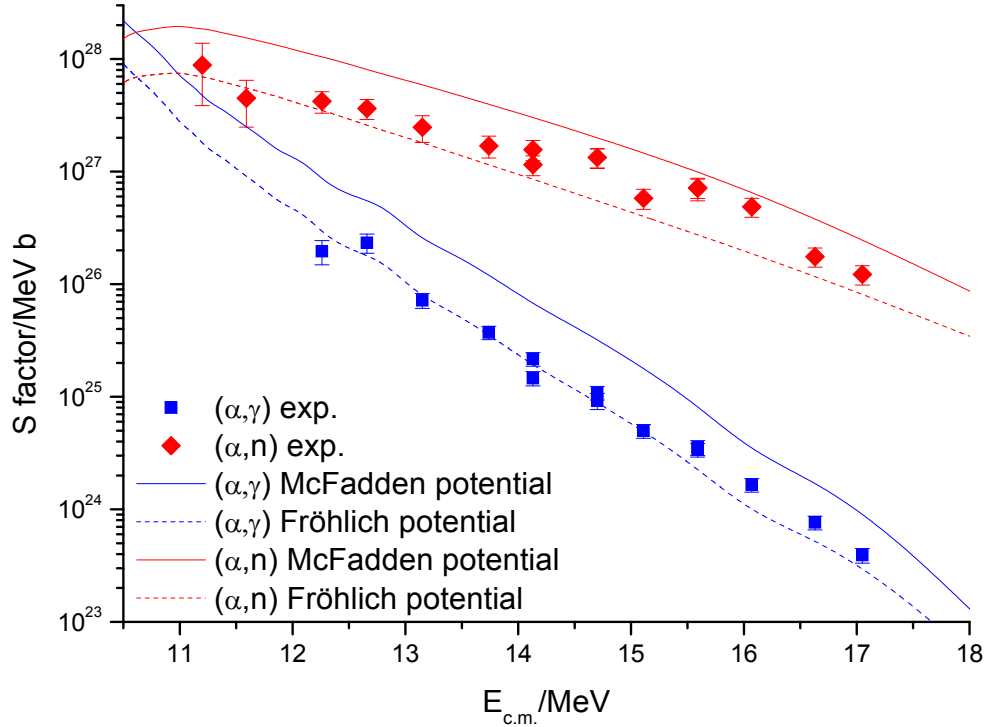}}}
\caption{\label{fig:results} (Color online) Measured and calculated cross section of the $^{151}$Eu($\alpha,\gamma)^{155}$Tb and $^{151}$Eu($\alpha$,n)$^{154}$Tb reactions (preliminary results). The two theoretical curves obtained with the NON-SMOKER$^\mathrm{WEB}$ code version 5.8w correspond to two different global $\alpha$-nucleus optical potentials of refs. \cite{mcf66} and \cite{fro02,rau03}.}
\end{figure}

Cross sections of the $^{151}$Eu($\alpha,\gamma)^{155}$Tb and $^{151}$Eu($\alpha$,n)$^{154}$Tb reactions have been measured in the center of mass energy range between 11.2 and 17.1\,MeV. At the lowest two energy points the yield from the ($\alpha,\gamma)$ reaction product was too low and therefore only the ($\alpha$,n) cross section could be obtained. The Gamow window at T\,=\,3\,GK temperature for this reaction is roughly between 8 and 12 MeV, so the measurements were carried out just above the astrophysically relevant energy range. In the measured energy range the cross section values are between about 5\,microbarns and 100\,millibarns. The final analysis of the data is still in progress. Figure \ref{fig:results} shows the preliminary results of both measured reaction in the form of the astrophysical S-factor. Also shown are the statistical model calculations performed with the  NON-SMOKER$^\mathrm{WEB}$ code version 5.8w \cite{NONSMOKER} using two different global $\alpha$-nucleus optical potentials. One is the standard potential of the NON-SMOKER$^\mathrm{WEB}$ code from McFadden and Satchler \cite{mcf66} and the other from Fr\"ohlich and Rauscher \cite{fro02,rau03}. The calculations reproduce well the energy dependence of the measured cross sections, while it seems that the absolute magnitude is better reproduced by the Fr\"ohlich and Rauscher potential. The final analysis and the interpretation of the data are in progress.

\section*{Acknowledgments}
This work was supported by ERC Grant 203175, GVOP-3.2.1.-2004-04-0402/3.0., OTKA (K68801, T49245), TUBITAK-Grant-108T508 and Kocaeli University BAP-Grant-2007/37.

\end{document}